# Influence of atomic tip structure on the intensity of inelastic tunneling spectroscopy data analyzed by combined scanning tunneling spectroscopy, force microscopy and density functional theory

Norio Okabayashi[1,2,*], Alexander Gustafsson[3], Angelo Peronio[1], Magnus Paulsson[3], Toyoko Arai[2], and Franz J. Giessibl[1]

[1]Institute of Experimental and Applied Physics, University of Regensburg, D-93053 Regensburg, Germany
[2]Graduate School of Natural Science and Technology, Kanazawa University, 920-1192 Ishikawa, Japan
[3]Department of Physics and Electrical engineering, Linnaeus University, 391 82 Kalmar, Sweden

* okabayashi@staff.kanazawa-u.ac.jp

Abstract
Achieving a high intensity in inelastic scanning tunneling spectroscopy (IETS) is important for precise measurements. The intensity of the IETS signal can vary up to a factor three for various tips without an apparent reason accessible by scanning tunneling microscopy (STM) alone. Here, we show that combining STM and IETS with atomic force microscopy enables carbon monoxide front atom identification, revealing that high IETS intensities for CO/Cu(111) are obtained for single atom tips, while the intensity drops sharply for multi-atom tips. Adsorbing the CO molecule on a Cu adatom [CO/Cu/Cu(111)] such that it is elevated over the substrate strongly diminishes the tip dependence of IETS intensity, showing that an elevated position channels most of the tunneling current through the CO molecule even for multi-atom tips, while a large fraction of the tunneling current bypasses the CO molecule in the case of CO/Cu(111).

Inelastic electron tunneling spectroscopy (IETS) with scanning tunneling microscopy (STM) is an effective method to analyze the vibrational modes of a single adsorbed molecule with sub-nanometer lateral resolution [1,2]. The vibrational energy of a molecule on a substrate strongly depends on the surrounding environment, such as the substrate structure and composition [3]. By studying these subtle changes of the vibrational energy using STM-IETS with a molecular functionalized tip, it has been demonstrated that STM-IETS can provide information on the inner structure of a molecule [4,5] similar to atomic force microscopy (AFM) [6]. These advantages of STM-IETS have accelerated research in related fields [7-16]. Owing to recent progress in the theoretical description of IETS [17-22], the qualitative understanding has been improved considerably, where the symmetry of the wave functions of a tip and a molecule on a substrate and a vibrational mode of the molecule is predicted to influence the efficiency of the inelastic process ($\gamma_{inel}$) for the tunneling current involving the molecule. In order to discuss $\gamma_{inel}$ from the intensity of IETS we have to consider that IETS intensity is described by the multiplication of the two factors: (1) the ratio of the tunneling current passing through a molecule to the total tunneling current ($I_{molecule}/I_{total}$) and (2) the efficiency of the inelastic process ($\gamma_{inel}$). These factors should in principle be affected by the geometrical structure of the substrate and the tip.

The geometrical structure of a metal tip apex can be determined by using carbon monoxide (CO) front atom identification (COFI) provided by AFM [23,24], where the tip apex of a force sensor is probed by a CO molecule that stands upright on a metal surface (inset of Fig. 1(e)). The metallic tip apex atom has a dipole moment induced by the Smoluchowski effect [25], whose direction is the same as that of the CO molecule [26]. Thus in the distance regime where the electrostatic interaction between the tip and the molecule dominates, the force between them is attractive. When the tip is scanned over the CO molecule, this attractive force appears as a dip (smaller value) in the frequency shift image for each atom at its apex, i.e., the number of the attractive force minima provides the number of atoms composing the tip apex [27].

In this paper, we have investigated the tip-structure dependent IETS for individual CO molecules on a Cu surface by combining STM and AFM. We have found that a tip with a single atom on its apex (single-atom tip) gives a stronger IET signal compared with a blunt tip consisting of four atoms on its apex (four-atom tip) for a CO molecule on a Cu(111) surface. However, the intensity between the two tips becomes comparable when a Cu adatom is inserted between the CO molecule and the Cu(111) substrate. From these findings, we will discuss the opposite electrode geometry dependent inelastic efficiency ($\gamma_{inel}$) and demonstrate the validity of the modern IETS theory.

The experiments are carried out in an ultra-high vacuum low-temperature (4.4 K) combined STM and AFM (LT-STM/AFM, Omicron Nanotechnology, Taunusstein, Germany). A Cu(111) surface was cleaned by repeated sputtering and annealing before adsorbing CO molecules on it. The force acting between a CO molecule and the apex of the metallic tip is measured by a qPlus sensor [28]. The sensor whose stiffness is $k$=1800 N/m oscillates at $f_0$=47375 Hz with a constant amplitude of 20 pm during all STM/AFM measurements. When an average force gradient $<k_{ts}>$ acts between the tip and the CO molecule, the sensor frequency is shifted by $\Delta f = f_0 <k_{ts}>/2k$ [29]. The current $<I_t>$ is averaged for many cycles of the sensor oscillation, since the bandwidth of the current amplifier is small compared to $f_0$. In order to measure the conductance ($dI/dV$) and IET signal ($d^2I/dV^2$), a modulation voltage (2338.7 Hz, 1 mV$_{rms}$) is added to the sample bias and the first and second harmonics in the current are detected by a lock-in amplifier (HF2LI, Zurich Instruments, Zürich, Switzerland). Throughout the whole text, the IET signal is normalized with the differential conductance, i.e., IETS=$(d^2I/dV^2)/(dI/dV)$ [10,14,18,20]. The tip is formed from an etched tungsten wire, cleaned by field evaporation and repeatedly poked into the Cu substrate to prepare various tip apexes [23,24,30]. The repeated poking processes probably cover the tip apex with Cu atoms. The poking processes also scatter Cu adatoms on the Cu(111) surface which are employed as another target by adsorbing a CO molecule and as an opposite electrode for the CO-functionalized tip [31].

Calculations of the IETS are performed with the density functional theory program Siesta [32]. From the relaxed geometries obtained by Siesta, elastic transport properties are calculated by attaching electrodes with TranSiesta [33]. Vibrational frequencies and IETS are obtained from the TranSiesta calculations using the post-processing package Inelastica [34]. The Siesta (TranSiesta) calculations utilize a supercell consisting of a 7 (17) layer thick 4×4 Cu slab together with the CO-molecule and a pyramidal tip modeled by 4 Cu atoms on the reverse side of the slab. The computations were performed using the following parameters: Perdew-Burke-Ernzerhof functional, 400 Ry real space cutoff, 4×4 k-points and a DZP (SZP) basis set for CO (Cu). The IETS calculations are then performed in the gamma-point.

Fig. 1(a)[(c)] shows a constant-height current image of a CO molecule on the Cu(111) surface obtained by a single-atom tip [a four-atom tip], as confirmed by the simultaneously acquired $\Delta f$ images (Fig. 1(b)[(d)]). For both tips we see the dip in the current image at the position of the CO molecule [35,36], where the current on top of the CO molecule is larger for the four-atom tip than for the single-atom tip owing to the larger tip area from which electrons can tunnel. Fig. 1(e) shows the IETS for CO molecules by the single-atom tip [37] and the four-atom tip, where identical current set-points are used for both tips. As described in the Appendix A, a background IETS measured on the copper surface is subtracted from that on the CO molecule. In the case of the single-atom tip, the frustrated translational (FT; ~4 meV) and frustrated rotational (FR; ~35 meV) modes of the CO molecule [7-9] are clearly seen in its IETS. However, the IETS intensity acquired with the four-atom tip is considerably smaller than that acquired with the single-atom tip: 65% decrease for both the FT and FR modes. A reduced IETS intensity is also observed for tips with two atoms and three atoms on its apex.

The strong intensity of the IETS provided by single-atom tips is confirmed by preparing different tips, which by COFI measurements (Fig. 2(a)) are single-atom tips. Cross-sections of constant-height current image are shown in Fig. 2(b) [38]. Note that tip #1 is the one used in Fig 1(a) and (b). In the case of tips #1 through #5, the minimum current acquired on the CO molecules is almost identical and the value is 24% of that on the Cu surface, thus these single-atom tips are judged to be sharp. The normalized IETS is also consistent for the different sharp single-atom tips (tips #1 to #5) (Fig. 2(c))(Appendix A). On the other hand, the IETS intensity with a single-atom tip having secondary-atoms outside of the apex which can contribute to the tunneling, is considerably weaker (Appendix A). This decrease originates from the increased fraction of tunneling electrons that do not interact with CO molecule and pass directly between the tip and the substrate. The decrease in IETS intensity for the four-atom tip can be similarly rationalized as a decreased ratio of tunneling current involving the CO molecule to the total current ($I_{CO}/I_{total}$). To investigate how the inelastic efficiency ($\gamma_{inel}$) depends on the geometry, we now present IETS measurements for a system where the tunneling current is dominantly passing through a CO molecule.

Figure 3(a) shows IETS for a CO molecule on a Cu adatom on the Cu(111) surface (Appendix A) with the single-atom tips [37] and the four-atom tip used in Fig. 1(e), where the current set-point is identical. In this case, the IETS intensity of the frustrated rotational (FR) mode with the four-atom tip is 20% smaller than the case of the single-atom tip, however, their overall intensity is comparable. Raising the vertical position of the CO molecule by one Cu adatom removes the direct tunneling channel between tip and substrate, causing almost all the current to pass through the CO molecule. The similarity of the intensity between the single-atom tips and four-atom tip indicates that $\gamma_{inel}$ does not depend on the tip electrode geometry investigated here. The same conclusion can be derived from the IETS with a CO-functionalized tip over (1) a Cu adatom and (2) the bare Cu(111) surface [39]. In the two situations investigated in Fig. 3(b), tunneling electrons are emitted from or injected to the single-atom on which the CO molecule is adsorbed, thus the electron beam is focused and $I_{CO}/I_{total}$ is expected to be large and similar. The similarity of the IETS intensity between two cases again indicates that the structure of the opposite electrode such as the Cu adatom and Cu plane does not strongly influence the $\gamma_{inel}$ for a CO molecule on the tip apex.

The role of the electrode opposite to a CO molecule in $\gamma_{inel}$ has been further investigated theoretically by using the TranSiesta code [33]. The computations utilize a localized basis set causing the calculated tunneling current to preferentially pass through the CO molecule regardless of the opposite electrode geometry [35], i.e., the direct tunneling between the tip and the substrate is underestimated. Thus the calculated IETS dominantly reflects the contribution of $\gamma_{inel}$ rather than that of $I_{CO}/I_{total}$. Fig. 4(a) shows the calculated IETS for a CO on a Cu (111) with the single- and three- atom tip. Here we can see that the intensity of IETS for the rotational mode is almost identical between two tips, which support our conclusion that the $\gamma_{inel}$ does not depend on the opposite electrode geometry.

The conclusion that $\gamma_{inel}$ does not depend on the opposite electrode geometry can be rationalized considering the symmetry of the tips states with respect to the molecular axis [18-20]. In the case of the CO molecule, the two-fold degenerate π symmetric molecular states dominantly contribute to the inelastic tunneling process (Fig. 4(b)), because these states are more localized on the O molecule than that of the σ symmetric state [19] (Appendix B). Taking into account that the FT and FR modes have a π symmetric character, the tip state with σ symmetry should effectively contribute to the inelastic tunneling process (Fig. 4(b)) (Appendix B). The relative contribution of the σ state to the total transmission is 50% for the three-atom tip and 57% for the single-atom tip, i.e., the contribution of the σ state drops about 12% from a single- to a three-atom tip, resulting in an almost identical $\gamma_{inel}$ in the calculation.

In contrast to the metallic tips investigated here, for the case of a CO functionalizes tip, the symmetry of the tip state is drastically changed from σ to π, which results in the inversion of the STM image contrast for a CO molecule on the Cu(111) surface from dip to bump [31,35]. This change of the tip state is predicted to decrease the efficiency of the IETS considerably for a CO molecule on a metal surface [19] in contrary to the present case.

In summary, by combining STM and AFM, we have investigated the dependence of the IETS intensity on the structure of the tip electrode for individual CO molecules for several Cu substrates. We have found that for the system where the current dominantly pass through the CO molecule by positioning this molecule on top of a Cu adatom, the IETS intensity is almost identical regardless of the tip geometry. This result indicates that the inelastic tunneling efficiency ($\gamma_{inel}$) is independent on the geometry of the tip electrode at least for a metallic tip. This conclusion demonstrates the validity of the modern IETS theory based on density functional theory and nonequilibrium Green's functions [18-20]. While we have found that single-atom tips provide a maximal IETS intensity and great reproducibility, single-atom tip are more reactive than multi-atom tips [23,24,40]. Therefore multi-atom tips may still be useful in cases where a high intensity is not key but a low perturbation to the vibrating molecule by the force field of a tip is desired.

We are deeply indebted to Thomas Frederiksen, Aran Garcia-Lekue and Alfred. J. Weymouth for stimulating discussions, to Daniel Meuer and Florian Pielmeier for the sample preparation and sensor construction and for Jascha Repp for various advices including a method to improve the resolution of IETS. This study was partially supported by a funding (SFB 689) from Deutsche Forschungsgemeinschaft (F.J.G); by JSPS "Strategic Young Researcher Overseas Visits Program for Accelerating Brain Circulation" (T.A. and N.O.); by a Grant-in-Aid for Young Scientists (B) (25790055) from MEXT (N.O.); by a grant from the Swedish Research Council (621-2010-3762) (A.G. and M. P.).

**Appendix A: Complete set of IETS with related STM images**

Figure 5 displays IETS data from five different tips. All these tips are single-atom tips, although none of them shows a COFI image that is perfectly symmetric with respect to rotations around the *z*-axis. Instead, the COFI images show slight asymmetries that could be attributed to a slight tilt of the plane of the second atomic layer of the tip. Nevertheless, the IETS spectra are essentially identical after background subtraction.

Decreased IETS for the single-atom, blunt tip is shown in Fig. 6 with the data by single-atom sharp tips, where the tip apex geometry is confirmed by COFI (Fig. 6(a)) and the sharpness of the tip apex is confirmed by the constant-height current measurement (Fig. 6(b)). When the tip is blunt, i.e., the tip constitutes a single-atom on its apex but has secondary-atoms outside of the apex (Fig. 6(c)), the IETS intensity is considerably decreased (Fig. 6(d)).

Complete set of the IETS for the CO molecule on Cu adatom and Cu substrate are shown in Fig. 7: IETS for (a) [(b)] CO/Cu(111) and (d) [(e)] CO/adatom with the single-atom tip #1 [#2], and for (c) CO/Cu(111) and (f) CO/Cu adatom with the four-atom tip. The topographic image of a CO on a Cu adatom is shown in Fig. 8 with the image of a CO and a Cu adatom on the Cu(111) surface.

**Appendix B: Detail of the theoretical IETS**

Theoretical IETS between two different set-points are shown in Fig. 9, where we see that the intrinsic IETS depends weakly on the tip apex geometry. Three most transmitting eigenchannels are shown in Fig. 10 for (a) the single-atom and (b) the three-atom tip, whose contribution to inelastic tunneling process for the FT and FR modes is summarized in table 1. We see that the contribution of the sigma state at the tip to the transmission is similar between the single-atom (57%) and three atom tip (50%), which results in the similar intrinsic IETS intensity. The calculated vibrational energies are summarized in table 2.

**References**

[1] W. Ho, J. Chem. Phys. 117, 11033 (2002).

[2] B. C. Stipe, M. A. Rezaei, and W. Ho, Science 280, 1732 (1998).

[3] H. J. Lee and W. Ho, Phys. Rev. B 61, R16347 (2000).

[4] C. L. Chiang, C. Xu, Z. Han, and W. Ho, Science 344, 885 (2014).

[5] P. Hapala, R. Temirov, F. S. Stefan Tautz, and P. Jelínek, Phys. Rev. Lett. 113, 226101 (2014).

[6] L. Gross, F. Mohn, N. Moll, P. Liljeroth, and G. Meyer, Science 325, 1110 (2009).

[7] L. J. Lauhon and W. Ho, Phys. Rev. B 60, R8525 (1999).

[8] A. J. Heinrich, C. P. Lutz, J. A. Gupta, and D. M. Eigler, Science 298, 1381 (2002).

[9] L. Vitali et al., Nano Lett. 10, 657 (2010).

[10] N. Okabayashi, M. Paulsson, H. Ueba, Y. Konda, and T. Komeda, Phys. Rev. Lett. 104, 077801 (2010).

[11] N. Okabayashi, M. Paulsson, H. Ueba, Y. Konda, and T. Komeda, Nano Lett. 10, 2950 (2010).

[12] N. Okabayashi, M. Paulsson, and T. Komeda, Prog. Surf. Sci. 88, 1 (2013).

[13] H. Gawronski and K. Morgenstern, Phys. Rev. B 89, 125420 (2014).

[14] K. J. Franke, G. Schulze, and J. I. Pascual, J Phys Chem Lett 1, 500 (2010).

[15] M. Grobis et al., Phys. Rev. Lett. 94, 136802 (2005).

[16] K. Motobayashi, Y. Kim, H. Ueba, and M. Kawai, Phys. Rev. Lett. 105, 076101 (2010).

[17] N. Lorente and M. Persson, Phys. Rev. Lett. 85, 2997 (2000).

[18] M. Paulsson, T. Frederiksen, H. Ueba, N. Lorente, and M. Brandbyge, Phys. Rev. Lett. 100, 226604 (2008).

[19] A. Garcia-Lekue, D. Sanchez-Portal, A. Arnau, and T. Frederiksen, Phys. Rev. B 83, 155417 (2011).

[20] E. T. R. Rossen, C. F. J. Flipse, and J. I. Cerda, Phys. Rev. B 87, 235412 (2013).

[21] G. Teobaldi, M. Penalba, A. Arnau, N. Lorente, and W. A. Hofer, Phys. Rev. B 76, 235407 (2007).

[22] A. Troisi and M. A. Ratner, Phys. Rev. B 72, 033408 (2005).

[23] J. Welker and F. J. Giessibl, Science 336, 444 (2012).

[24] T. Hofmann, F. Pielmeier, and F. J. Giessibl, Phys. Rev. Lett. 112, 066101 (2014).

[25] R. Smoluchowski, Phys. Rev. 60, 661 (1941).

[26] M. Schneiderbauer, M. Emmrich, A. J. Weymouth, and F. J. Giessibl, Phys. Rev. Lett.

112, 166102 (2014).

[27] M. Emmrich et al., Science 348, 308 (2015).

[28] F. J. Giessibl, Appl. Phys. Lett. 76, 1470 (2000).

[29] F. J. Giessibl, Phys. Rev. B 56, 16010 (1997).

[30] M. Emmrich et al., Phys. Rev. Lett. 114. 146101 (2015).

[31] L. Bartels, G. Meyer, and K. H. Rieder, Appl. Phys. Lett. 71, 213 (1997).

[32] J. M. Soler et al., J. Phys.: Condens. Matter 14, 2745 (2002),

[33] M. Brandbyge, J. L. Mozos, P. Ordejón, J. Taylor, and K. Stokbro, Phys. Rev. B 65, 165401 (2002).

[34] T. Frederiksen, M. Paulsson, M. Brandbyge, and A. P. Jauho, Phys. Rev. B 75, 205413 (2007).

[35] A. Gustafsson and M. Paulsson, arXiv:1512.00702v1.

[36] R. K. Tiwari, D. M. Otalvaro, C. Joachim, and M. Saeys, Surf. Sci. 603, 3286 (2009).

[37] The data is an average of IETS by the tip in Fig.1(b) (tip #1 in Fig. 2(a)) and the tip #2 in Fig. 2(a), since these two tips are adopted for IETS measurements of CO/Cu/Cu(111) in addition to CO/Cu(111).

-

[38] The current on the Cu(111) surface is slightly modulated depending on the substrate position owing to the standing waves. The averaged current $<I_t>$ in Fig. 2(b) is normalized such that the current on the leftmost position is 1.5 nA.

[39] The tip before the CO functionalization has a single-atom on its apex, and is sharp in the sense of Fig. 2.

[40] T. Hofmann, F. Pielmeier, and F. J. Giessibl, Phys. Rev. Lett. 115, 109901 (2015).

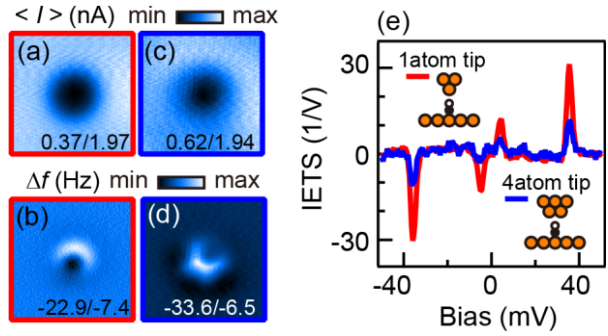

Fig. 1. (color online) Constant-height, (a)[(c)] current and (b)[(d)] frequency shift images (1.5 nm×1.5 nm) for a CO molecule adsorbed on Cu(111) by a single-atom tip [four-atom tip]. The tip height is set on the Cu(111) substrate at a sample bias $V_t=-1$ mV and an average current $<I_t>$=1.5 nA. (e) Normalized IETS for a CO molecule at a set-point of $V_t=-50$ mV and $<I_t>$=5 nA, where the IETS on the Cu(111) surface is subtracted.

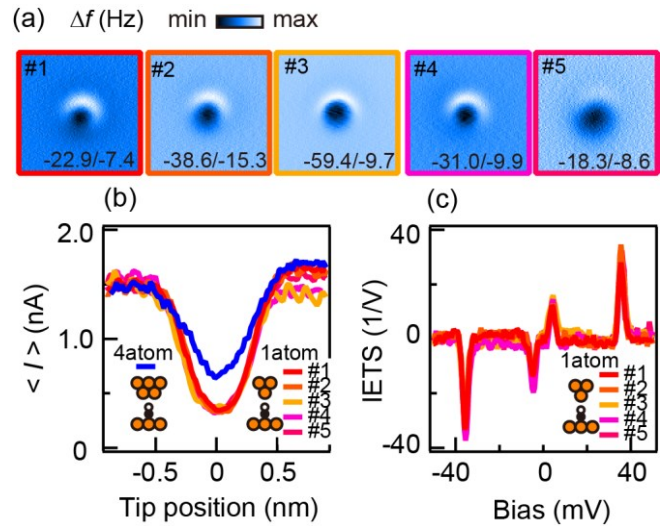

Fig. 2. (color online) (a) COFI images (1.5nm×1.5nm) and (b) cross-sections of the constant-height current images on a CO molecule with different single-atom (tip #1 to #5) tips at the set-point of $V_t=-1$ mV and $<I_t>$=1.5 nA on the Cu(111) surface. As a reference, the cross-section of the current image with the four-atom tip in Fig. 1(c) is added in (b). (c) IETS for CO molecules with the same tips where the set-point on a CO molecule is $V_t=-50$ mV and $<I_t>$=5 nA.

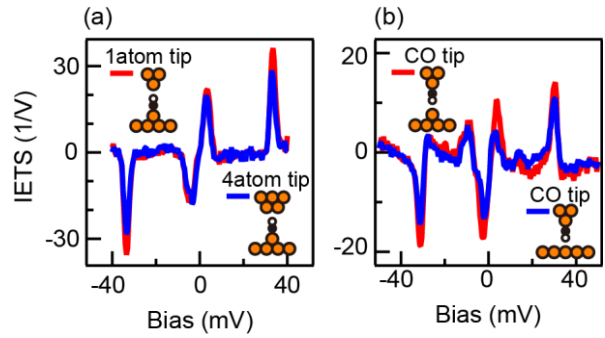

Fig. 3. (color online) (a) IETS with the single-atom tips (red) and the four-atom tip (blue) for a CO on a Cu adatom. (b) IETS of a CO-functionalized tip for a Cu adatom (red) and the bare Cu(111) surface (blue). In both cases [(a) and (b)], the tip-height is set at $V_t=-50$ mV and $<I_t>=5$ nA on the measurement points.

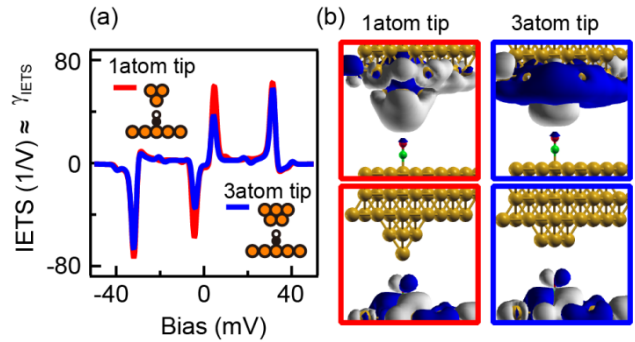

Fig. 4. (color online) (a) Calculated IETS with a single-atom tip (red) and three-atom tip (blue) for a CO on a Cu(111), where the tips is positioned such that the calculated elastic current is nearly identical ($V_t=-50$ mV and $I_t\approx 5$ nA). Broadening by the modulation voltage (1.0 mV$_{rms}$) and temperature (4.4 K) is included. (b) Most important tip (upper panels) and molecular (lower panels) scattering states for the inelastic scattering by the rotational and translational modes.

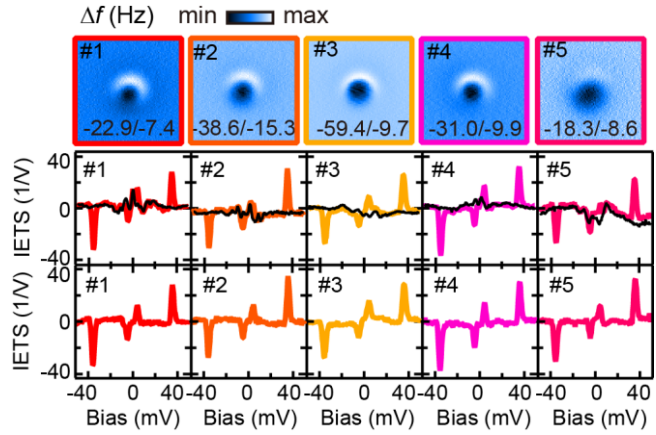

Fig. 5 (color online) Reproducibility of IETS spectra for five different tips #1 to #5. The data is the same as that shown in Fig. 2(c), but we display the spectra individually including background subtraction here. The top row shows the constant height COFI (carbon monoxide front atom identification) frequency shift profiles (in Hz) for the five different tips. The center row displays the IETS signal in color and the background spectra in black. The bottom row displays the net IETS signal without background.

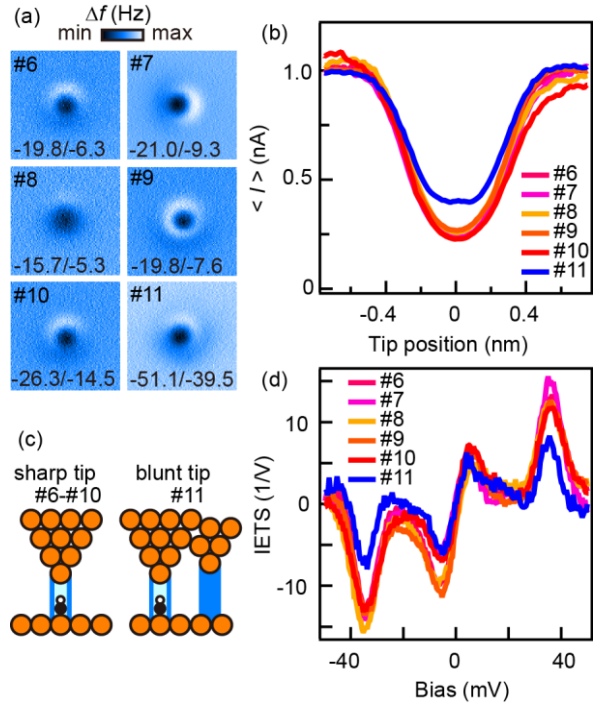

Fig. 6. (color online) (a) COFI images (1.5nm×1.5nm) and cross-sections of the constant-height current images for a CO molecule with various single-atom tips (tip #6 to #11). The set-point on the Cu(111) surface is $V_t=-1$ mV and $<I_t>=1.5$ nA for the COFI measurements and $V_t=-1$ mV and $<I_t>=1$ nA for current measurements. (c) Schematic images of the single-atom sharp and blunt tips. (d) IETS for CO molecules with the tip #6 to #11 where the set-point on a CO molecule is $V_t=-50$ mV and $<I_t>=10$ nA. Note that the modulation voltage used here is 3.5 mV$_{rms}$, which is larger than the value adopted for the case of the main text (1 mV$_{rms}$).

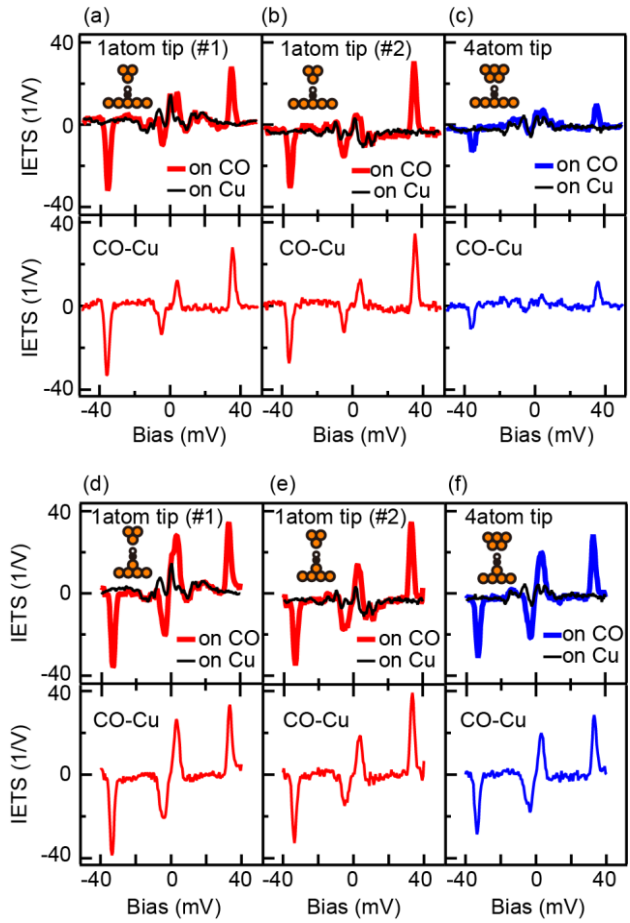

Fig. 7. (color online) Set of IETS on the CO molecule and the Cu(111) substrate (upper panel) and after the background subtraction (lower panel). Experimental conditions: the single-atom tip (tip #1[#2]) to (a)[(b)] CO/Cu(111) and (d)[(e)] CO/Cu/Cu(111); the four-atom tip to (c) CO/Cu(111) and (f) CO/Cu/Cu(111). For the background measurements, IETS on the Cu(111) surface are measured and averaged for 16 different points. A set-point of $V_t=-50$ mV and $<I_t>=5$ nA on the CO molecule has been used for all measurements.

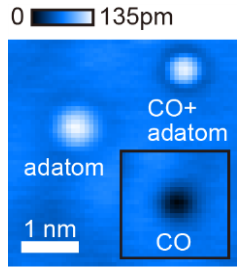

Fig. 8. (color online) Constant-current images of a Cu adatom and a CO molecule on a Cu adatom by a single-atom tip ($V_t=-10$ mV, $<I_t>=100$ pA). The inset shows a CO molecule on the Cu(111) surface.

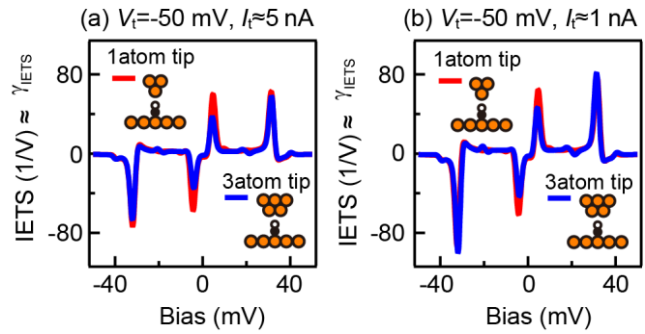

Fig. 9. (color online) Calculated IETS data between a single-atom tip and a three-atom tip are displayed for two different set-points where the transmission acquired by TranSiesta is almost identical between the single-atom- and three-atom tip: $V_t=-50$ mV, $I_t\approx 5$ nA and $V_t=-50$ mV, $I_t\approx 1$ nA. Note that Fig. 4(a) in the manuscript shows the data for the former set-point.

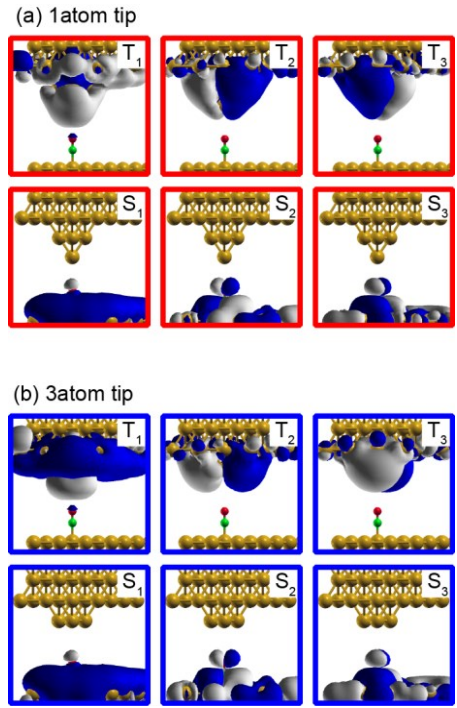

Fig. 10. (color online) Isosurface plot of the three most transmitting eigenchannels between the CO molecule on the Cu(111) surface and (a) the single-atom tip [(b) the three-atom tip] at a set-point of $V_t=-50$ mV, $I_t\approx 5$ nA. The top (bottom) row of scattering states originates from the tip (substrate) side.

Table. 1. Average (over two degenerated vibration) partial contribution (%) of each scattering state (see Fig. 9) in the inelastic process of the FT and FR modes with the elastic transmission for the single-atom and three-atom tips at a set-point of $V_t=-50$ mV, $I_t\approx 5$ nA.

|  | 1 atom tip | T1($\sigma$) | T2($\pi$) | T3($\pi$) |
|---|---|---|---|---|
| FT (%) | S1($\sigma$) | 0.19 | 1.57 | 1.63 |
|  | S2($\pi$) | 46.65 | 0.01 | 0.10 |
|  | S3($\pi$) | 44.64 | 0.01 | 0.07 |
| FR (%) | S1($\sigma$) | 0.27 | 2.00 | 1.95 |
|  | S2($\pi$) | 48.00 | 0.02 | 0.12 |
|  | S3($\pi$) | 46.32 | 0.02 | 0.07 |
| Transmission ×10$^{-4}$ |  | 9.00 | 3.37 | 3.29 |
| Transmission (%) |  | 57 | 22 | 21 |

|  | 3 atom tip | T1($\sigma$) | T2($\pi$) | T3($\pi$) |
|---|---|---|---|---|
| FT (%) | S1($\sigma$) | 0.09 | 3.36 | 3.33 |
|  | S2($\pi$) | 45.17 | 0.16 | 0.08 |
|  | S3($\pi$) | 43.78 | 0.08 | 0.15 |
| FR (%) | S1($\sigma$) | 0.16 | 6.22 | 6.53 |
|  | S2($\pi$) | 37.63 | 0.11 | 0.20 |
|  | S3($\pi$) | 35.37 | 0.31 | 0.85 |
| Transmission ×10$^{-4}$ |  | 8.49 | 4.27 | 4.20 |
| Transmission (%) |  | 50 | 25 | 25 |

Table. 2. Vibrational energy (meV) of FT and FR modes at a set-point of $V_t=-50$ mV, $I_t\approx 5$ nA, which are calculated by allowing the CO molecule and the one (three) Cu atoms of the 1 (3)-atom tip to move (dynamical region).

|  | FT1 | FT2 | FR1 | FR2 |
|---|---|---|---|---|
| 1atom | 4.59 | 4.61 | 31.6 | 31.8 |
| 3atom | 4.25 | 4.28 | 31.7 | 31.9 |